# Vigilance Overload Measured by Computerized Mackworth Clock Test

Ipek Ustun, Ege Ozer, Erim Habib, Burcin Tatliesme, Ata Akin

***Abstract*—** This paper studied the change of vigilance based on stimulus coming consecutively using the computerized version of the Mackworth Clock Test run from *PsyToolkit* website. 7 participants (16.57 ±1 years old, 2 males), performed 10 consecutive trials in order to measure whether or not it is a realistic goal for high school students to display the level of vigilance expected from them in class. Success percentages were calculated by dividing the number of correct jumps to the total number of jumps. The results indicated that while the average success percentage for all subjects remained relatively stable over the 10 trials (79% ± 7%), success percentages drop relatively as the number of jumps increase. Success rate dropped from 90% (2 jumps) to 70% (7 jumps). We conclude that there is an upper limit of vigilance that should be expected from students when they are exposed to more than 4 randomly occurring attention requiring task within a minute.

## İ. Introduction

THE Mackworth clock was originally created by Norman Mackworth in order to simulate the long-term monitoring done by radar operators of the British Air Force during World War II. The Mackworth clock is still used today in vigilance research in various forms, including computerized versions of the test. Vigilance, or sustained concentration, is defined as the ability to maintain concentrated attention over prolonged periods of time [1].

The first applications of the test usually lasted usually for two hours and were performed manually. Norman Mackworth's primary goal of using the test on human subjects was to observe the effect of stress factor on working class and their performance. The subject was asked to do the test for two hours and his performance was compared with one-hour performance. The results showed that the efficiency of the subject who performed the Mackworth task for two hours decreased more rapidly than the subject who performed it for an hour. This research addressed the phenomenon that vigilance would decrease with the over-repetition of the task for a long time [2].

Mackworth clock test has also been used in order to investigate sleep deprivation. Williamson *et al.* recruited 39 subjects with performing Mackworth clock test and other types of vigilance-measuring tests. Results showed that the performance of subjects who had been awake between 17-19 hours was equally low as the time when subjects were supplied with 0.1% alcohol dosage. [3] The task was also used to investigate the effect of caffeine on sleep deprivation. McIntire *et al.* compared the effect of caffeine and transcranial direct current stimulation (tDCS). Results showed that caffeine did not have a significant effect on subject's performance as they were performing the Mackworth test. The performance on Mackworth clock test revealed that caffeine did not significantly change the success of subjects. [4]

The aim of this experiment was to measure vigilance in high school students aged 15-18 using the Mackworth clock test. High school students are expected to concentrate for long stretches of time during classes, and a computerized version of the Mackworth clock test was employed in order to measure whether or not it is a realistic goal for high school students to display the level of vigilance expected from them. It was predicted that, since the number of jumps was randomized, there would be a decrease in the subjects' performance consistent with the increase in number of jumps. The effect of number of jumps in one test episode to success rate was also investigated to understand the relationship between over stimulation to vigilance.

## İİ. Methods

In this experiment, the data which was based on correct and wrong jumps of jumps were collected from 7 high school students (Mean age =16.57, ±STD = 0.9759, range 15-18 years old, 2 males) who did 10 trials in total without having any break. Subjects were placed in a silent room in order to maintain the stability of controlled factors and prevent the effect of outside factors. They were sitting in an upright position and 30 cm far from the screen.

In this Mackworth Clock Test which was run from *PsyToolkit* website, the subjects were told to watch the clock hand moving around and when they saw a double jump they were tasked to press the space bar immediately (within 1 second). The subjects were given feedback immediately whether their choice was right or wrong. If they pressed the spacebar when there was no double jumping of the clock hand or if they failed to detect it, they got an error signal (red light) as seen in Fig. 1 (a) and when they pressed the space bar correctly they were getting a feedback (green light) as seen in Fig. 1 (b).

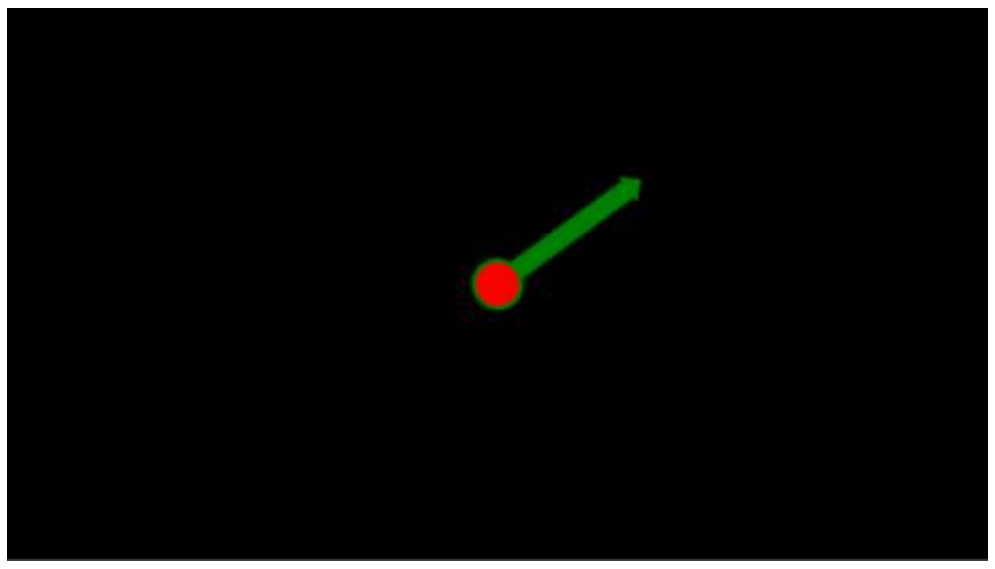

(a)

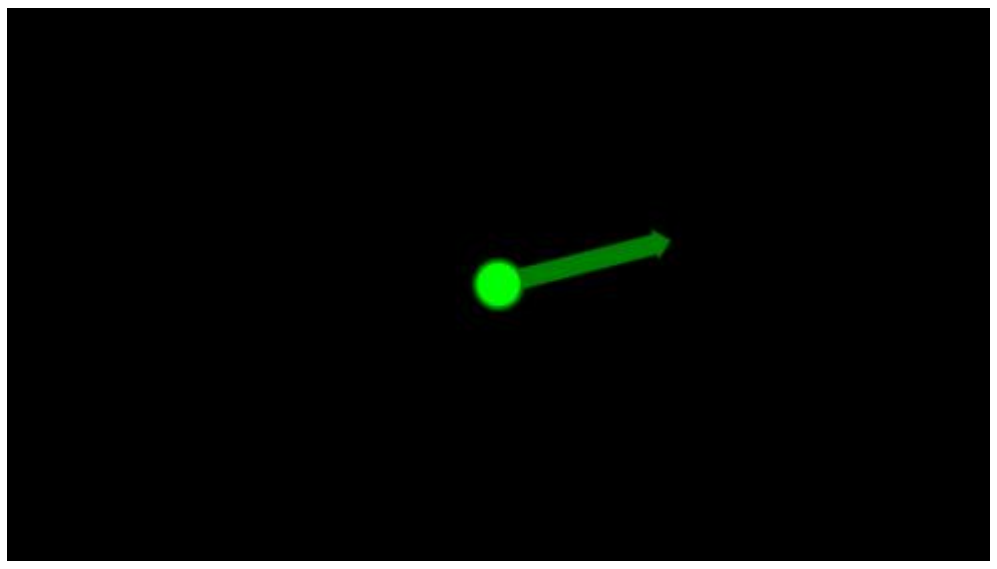

(b)

**Figure 1 Screen shot of the Macworth Test where (a): feedback screen when missed or wrongly pressed (b): feedback screen when correctly pressed**

But the number of jumps were set randomly by the system's algorithm before each trial and the amount of jumps may vary in each trial. 2-10 jumps were presented at different positions around the of the clock [5]. At the end, results were collected from each subject separately. After the collection of data, success percentages were calculated and plotted on a graph by putting correct jumps over total jumps into percentages.

## III. RESULTS

We first investigated how the performance varied over 10 trials. Figure below shows the average of success percentages of 6 subjects over the course of ten trials.

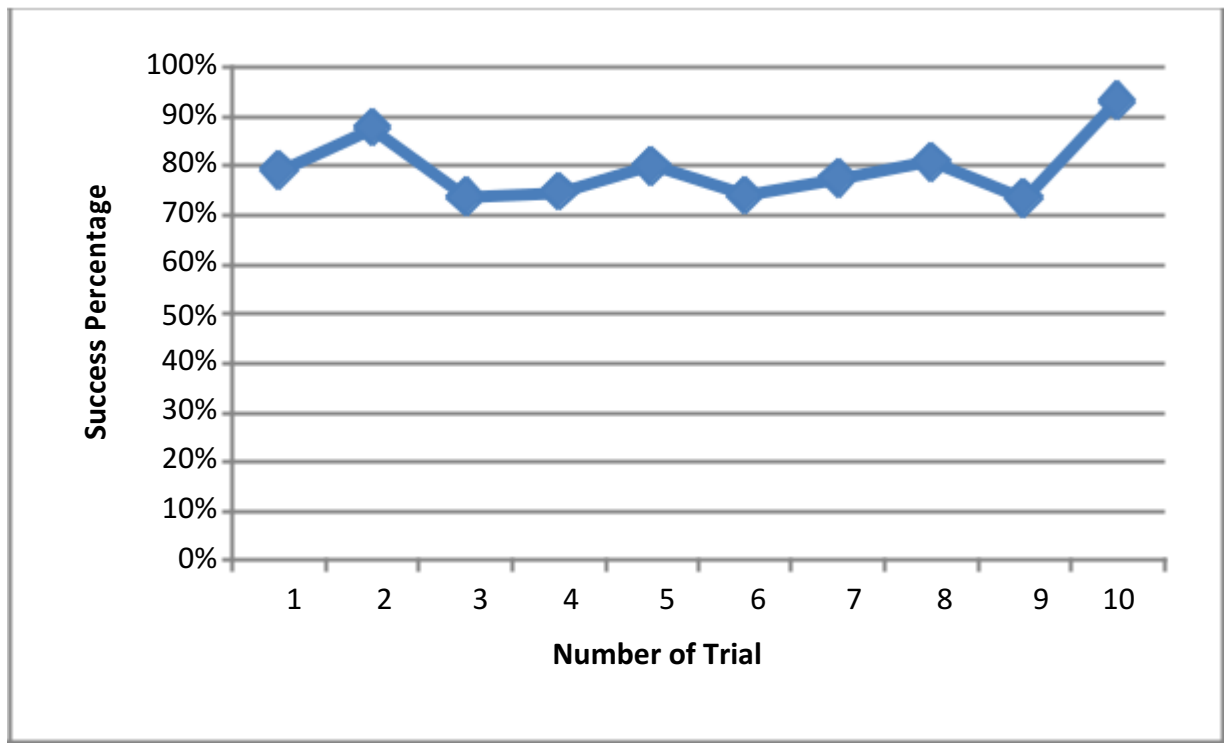

**Figure 2 Average success percentage over the 10 trials**

As seen in Figure 2, the variability is low, with an average of 79% and a standard deviation of 7%. There is no statistical difference between the first and last performance values (p=0.156). Similarly, no significant difference was observed between the highest performance (second trial, mean accuracy: 88%) and lowest performance (9$^{th}$ trial, mean accuracy: 73%), p=0.0118.

We then analyzed the findings with respect to number of jumps, rather than number of trials. Figure 3(a) shows when the data are rearranged and plotted with respect to 2 jumps through 10 jumps in total. In the first graph a decrease is observed in the overall success percentage. Only three people were presented with jumps greater than 8 jumps. Regarding the fact that there is an inadequacy in the number of data, last two jumps are discarded and success percentage of 2 through 8 jumps are plotted in order to better interpret the results as seen in Figure 3 (b).

In **Figure 3** (b), a fluctuation and a decrease in success percentage were observed. The average accuracy of success percentage of 7 participants was 82,43%. When 4 jumps were made in the test, subjects achieved a peak success. Between 2 and 4 jumps an increase in success percentages was observed as 83.33% in 2 jumps, 86.66% in 3 jumps and 94.17% in 4 jumps. However, after 4 jumps there was a decline in the success percentages until 6 jumps. Between 6 and 7 jumps an increase by 3.51% was observed in the graph and also between 7 and 8 jumps a decrease in success percentage by 14.62% was observed. Between 2 and 7 jumps, overall success percentage was between 77.78% and 94.17%. However, after 7 jumps success percentages are below 75% in both figures. Between 3-4-5 and 6-7-8 jumps, a t-test was applied and gave a p-value of 0.0930427. Between 3 jumps and 8 jumps a t-test was made and its p-value was 0.4012829. Between 4 jumps and 6 jumps a t-test is made which is equivalent to 0.0708638.

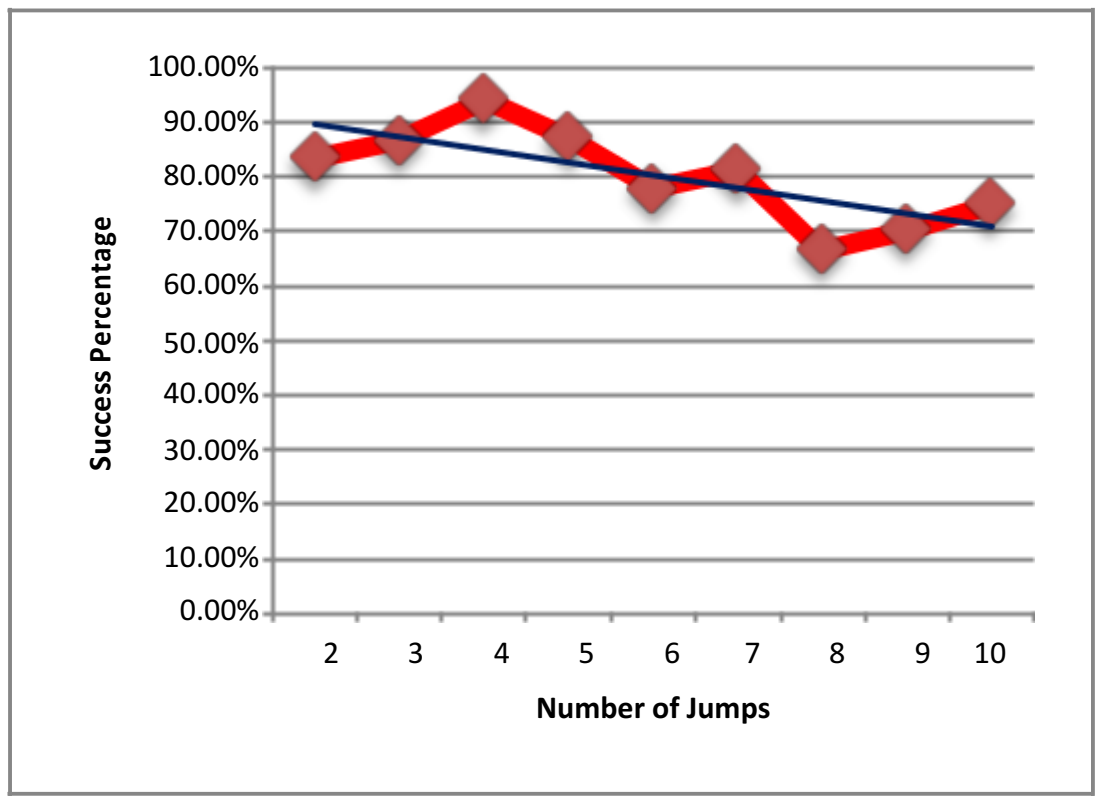

(a)

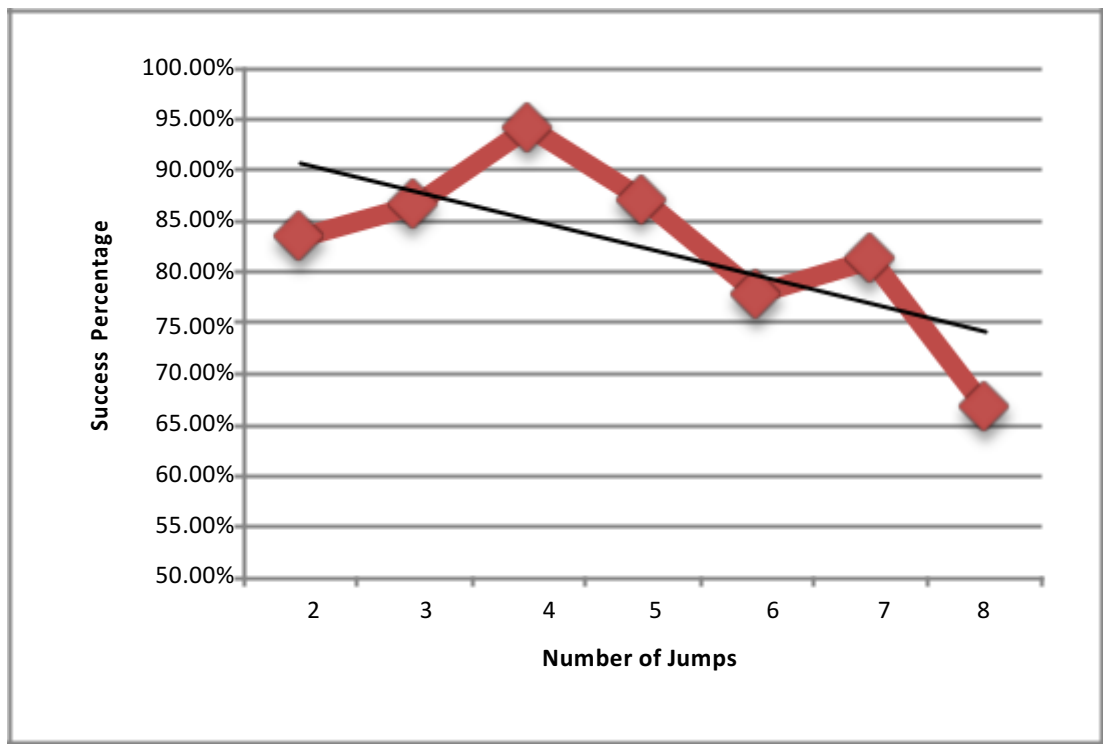

(b)

**Figure 3 (a)** Mean of success percentages for all the jumps (b): Mean of success percentages for jumps 2 through 8.

The standard deviations are also calculated and documented into a graph as shown in Figure 4. At 2 jumps, there is a peak in the graph where there are only 3 values. After 2 jumps, there is a linear fall till 4 jumps. In 3 and 4 jumps, there are 5 values for each of them. After 4 jumps, there is an increase in standard deviation, at 7 a fall, and at 8 jumps there is again an increase in standard deviation which is substantially lower than the standard deviation at 2 jumps.

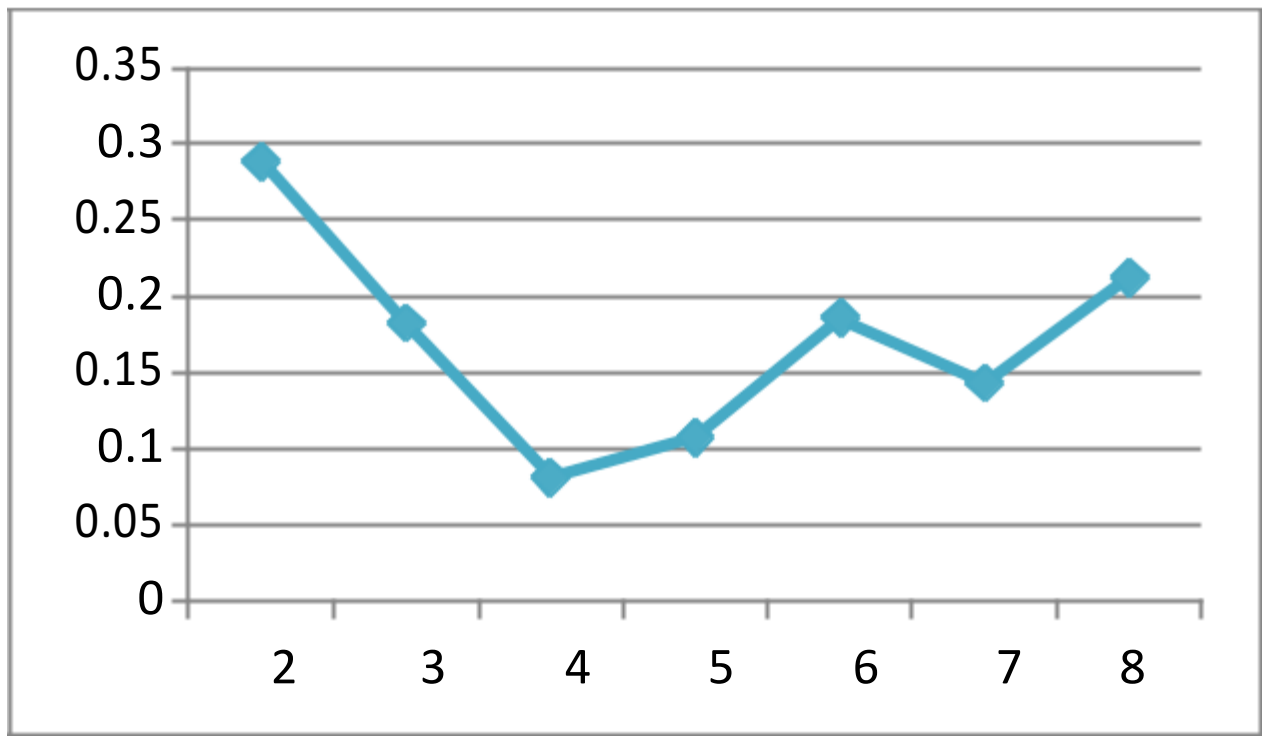

Figure 4 Standard Deviation of success percentages

TABLE I
Mean and Standard Deviations of Success Percentages According to The Number of Jumps

| | Mean | STD |
|---|---|---|
| 2 Jumps | 83% | 29% |
| 3 Jumps | 86% | 18% |
| 4 Jumps | 94% | 8% |
| 5 Jumps | 87% | 10% |
| 6 Jumps | 77% | 18% |
| 7 Jumps | 81% | 14% |
| 8 Jumps | 66% | 21% |

## IV. Discussion

The version of Mackworth Clock Test that is used in this experiment is considerably different from the original version. In the original Mackworth's experiment, in every 30 minutes there were 12 jumps repeated and there was no break between half-hour periods. The jumps were placed in the following intervals: 0.75 min, 0.75 min, 1.50 min, 2 min, 2 min, 1 min, 5 min, 1 min, 1 min, 2 min, and 3 min; 20 minutes in total. Mackworth did not report when the first jump occurred, but he indicated that there were not any jumps during the last 10 min of each half hour period [6]. But in the computerized version used in this study, the jumps were given randomly by the algorithm. There was not any accurate number of jumps or time intervals. That is why, in this experiment subjects did not have the same number of jumps in every trial.

In Mackworth's experiment, participants' task was to press a Morse key when they see a jump [6] which is same in this experiment. However, Morse's experiment tasked the subjects to press the key within 8 seconds [6] whereas in our experiment, subjects were given 1 second. Mackworth only reported missed jumps (failures to respond to a signal within 8 s), which were calculated by subtracting successfully detected jumps from total jumps [6]. In our study, the number of correct jumps were divided by total jumps and put into percentages. Mackworth did not provide any standard deviations. Focusing on Mackworth's data, performance, as measured by misses, significantly decreased from the first half hour to the second [6]. Also in our results, there was a decreasing trend. This is because when the number of jumps increases, subjects struggle more to detect the jumps in the same time given.

Till 4 jumps, there was an increase in overall success percentage. At 4 jumps, subjects achieve a peak success rate. After 4 jumps a decrease in overall success percentage was observed except the fluctuation at 7 jumps. Decrease in success percentage after 4 jumps is probably caused by a decline in subjects' vigilance as the number of jumps increased. The increase in success percentage until 4 jumps may be caused by the jump intervals. As the number of jumps increases until 4 jumps, vigilance was probably increasing. After 4 jumps, jump intervals shortened which is the reason for the decline in success percentage. As showed in Fig. 1 and Fig. 2, a Mackworth Clock Test containing 4 jumps is the best version to successfully measure the drop-in subjects' vigilance.

The decrease in standard deviation until 4 jumps may be caused by the differences in number of data collected from each subject. Regarding the fact that the experiment was made only on 7 subjects and the jump number in the test was given randomly, there were slight differences in the number of collected data. In order to increase the reliability of results, the data after 8 jumps are trimmed. But this does not mean that the remaining jumps have all the same amount of data. The high fluctuation in the standard deviation of 2 and 3 jumps is probably caused by the sudden difference in amounts of data between two consequent jumps. At 2 jumps, there are 3 data while at 3 jumps there are 5.

Each subject made 10 trials in the test because it is believed that at the end of 10 trials subjects would be bored and this state of boredom will cause an unexpected drop in success percentages which will prevent us from observing the effect of number of jumps to the results.

## V. Limitations and Future Studies

### A. Limitations

The test used in this study, the computerized version of Mackworth Clock Test, was taken from PsyToolkit. The algorithm of this test was programmed to give jumps randomly. Thus, the algorithm did not send a previously set number of jumps in each trial. This is why in each trial subjects faced different numbers of jumps and this presented different amount of data in each detection number.

### B. Improvements for Future Studies

Although the obtained results support our hypothesis, many improvements can be made in order to enhance the reliability of the results. In future studies, the number of subjects can be expanded. Also, different age groups may be constituted in order to analyze the differences between different ages. Last but not least, a Mackworth Clock Test which sends specific number of jumps would increase the reliability of the data collected.

## VI. Conclusion

In this experiment, it was observed that the increase in the number of jumps lead to a noticeable decrease in the subjects' performance as predicted in the hypothesis. The decrease in subjects' vigilance over time and with the increase in the number of jumps indicates that it gets harder for high school students to concentrate over long stretches of time.

## Acknowledgment

Authors would like to thank the subjects that participated in this study.